\documentclass[traditabstract]{aa}
\usepackage{graphicx}                    
\usepackage{bm}
\usepackage{amsmath}
\usepackage{txfonts}                    
\usepackage{natbib}                     
\bibpunct{(}{)}{;}{a}{}{,}
 \usepackage{threeparttable}
 \usepackage{amssymb}                    

\renewcommand{\vec}[1]{\mbox{\boldmath $#1$}}
\def \i {\ensuremath{\rm{i}}}
\renewcommand{\vec}[1]{\mbox{\boldmath $#1$}}
\def \d{{\rm d}}

\def \Om  {{\it \Omega}}
\def \Omin  {{\it \Om_{\rm in}}}
\def \Omout  {{\it \Om_{\rm out}}}

\def \rin {r_{\rm in}}
\def \Rin {R_{\rm in}}
\def \Rout {R_{\rm out}}
\def \Rey {\ensuremath{\rm{Re}}}
\def \Ha {\ensuremath{\rm{Ha}}}

\def \Pm {\ensuremath{\rm{Pm}}}

\def \Ra {\ensuremath{\rm{Ra}}}
\def \E {\ensuremath{\rm{E}}}
\def \Mm {\ensuremath{\rm{Mm}}}

\def \etaT  {\ensuremath{\eta_{\rm T}}}
\def \nuT  {\ensuremath{\nu_{\rm T}}}

\def\beg{\begin{equation}}
\def\ende{\end{equation}}
\newcommand{\gsim}{\lower.7ex\hbox{$\;\stackrel{\textstyle>}{\sim}\;$}}
\newcommand{\lsim}{\lower.7ex\hbox{$\;\stackrel{\textstyle<}{\sim}\;$}}
\def\curl{{\rm curl}} 
\def \h{{B_{\rm max}}}

\def\aap{Astronomy \& Astrophysics}

\def\ara\&a{Ann. Rev. Astronomy Astrophysics}

\def\apjs{Astrophys. J. Suppl.}

\begin{document}



\title{Tayler instability and dynamo action  of  cylindric  magnetic rings}
\titlerunning{Tayler instability and dynamo action}
%
   \author{G.~R\"udiger\inst{1,2} \and M.~Schultz\inst{1}}


%
  \institute{Leibniz-Institut f\"ur Astrophysik Potsdam (AIP), An der Sternwarte 16, D-14482 Potsdam, Germany,
 \and
  University of Potsdam, Institute of Physics and Astronomy, D-14476 Potsdam, Germany
}


\date{Received; accepted}
 
\abstract{The Tayler instability of an azimuthal magnetic field with one or two  ``rings'' along the radius is studied for an axially unbounded Taylor-Couette flow. The rotation law of the conducting fluid is a quasi-Keplerian one. Without rotation  all toroidal fields are  the more  destabilized the more the  radial profiles differ from the uniformity. For medium Reynolds numbers of rotation, however,  the behaviour of the lines of neutral stability strongly depend on the magnetic Prandtl number. For $\Pm=1$ the differential rotation  matches the  instability lines of azimuthal fields with and without  rings so that the maximally possible Reynolds numbers for  fields with smooth radial profiles and such with rings do hardly differ. The magnetic Mach number of the considered examples  are of the astrophysically relevant  order of magnitude,  $10<\Mm<20$.

The nonaxisymmetric instability fluctuations form a weak $\alpha$ effect of the mean-field electrodynamics which always changes its sign  between the walls independent of the Reynolds number, magnetic Prandtl number or the radial profile of the magnetic background field. The resulting dynamo  modes   work on a similar  axial scale as the Tayler instability, hence they are  {\em small-scale} dynamos. The fields are axially drifting with high phase velocity  where at certain periods the azimuthal fields with one-ring geometry develop to fields with two rings along the radius. 
It is still open whether and how a nonlinear dynamo model may overcome this puzzling complication.  
}


\maketitle
%
\section{Introduction} \label{Section1}
One of the most challenging problems in astrophysical fluid
dynamics is the investigation of the influence of the global rotation
on the  stability of  toroidal magnetic fields. The condition
\beg
\frac{\d}{\d R} (R B_\phi^2) \leq 0
\label{Tayler}
\ende
is  sufficient and necessary  for stability of a stationary ideal magnetized fluid against nonaxisymmetric perturbations \citep{T73}. As a consequence, one finds uniform or outwardly increasing magnetic fields unstable against nonaxisymmetric perturbations. This is in particular   true for the toroidal field with  $B_\phi\propto R$ produced by   uniform electric currents. The existence of the nonaxisymmetric instability for such a nonrotating  `$z$-pinch' has experimentally been shown by \cite{SS12} using  liquid GaInSn as the conducting fluid penetrated by an axial electric current.

Flows  of the Chandrasekhar-type,  where the background field and the background flow  have identical  isolines,   are unstable against nonaxisymmetric perturbations if at least one of the diffusivities  is non-zero. For $\Pm\ll 1$ the onset of the instability   also scales with the Reynolds number  and the Hartmann number, i.e. the neutral stability curves converge for $\Pm\to 0$ in the Hartmann number/Reynolds number plane.  A prominent example of this class of magnetohydrodynamic flows is the axially unbounded rigidly rotating $z$-pinch exhibiting  toroidal  flows and fields which only vary with   the distance $R$ from the rotation axis \citep{GR16}.

However, for experimental and astrophysical applications of the instability theory the electric current will never be uniform (as it is necessary for the pinch-type solutions) or, with other words, the toroidal magnetic field profile will never be monotonously increasing. It will be more realistic to probe the stability/instability of toroidal fields having a maximum somewhere between the boundaries. For simplicity we shall only consider the maximum in the radial center  of the gap. 
The ring is  added to a quasi-uniform background field. 
Throughout the whole paper we have used a quasi-Keplerian rotation law where the cylinders´ rotation rates fulfill  the Kepler law. We shall see that without rotation the toroidal fields with  a ring-shaped structure are more  unstable against nonaxisymmetric disturbances than a quasi-uniform toroidal field  between the cylinders, see also \cite{DG09} for axisymmetric instability. This particular result does not depend on the magnetic Prandtl number. The opposite is true for models with differential rotation. Then the  rings do not play an important role for magnetic Prandtl number unity but for small magnetic Prandtl number they do. 

Another interesting question concerns the construction of dynamos from the interaction of magnetic instabilities and differential rotation \citep{S02}. This is indeed possible. Guseva et al. have shown dynamo action by numerical simulations of a quasi-Keplerian Taylor-Couette flow  including a   toroidal magnetic field which only depends on the radius in form of  $1/R$  \citep{GH17}. There is no axial large-scale current so that the field is stable without rotation. With differential rotation, however, this very simple system  may become unstable against nonaxisymmetric perturbances for suitable choices of parameters, \citep{RG14}. It is shown by simulations   that   toroidal field is maintained after switching off its external sources  if  $\Pm=10$ rather than  for $\Pm=1$. The axial scales of the field are similar to the radial scale of the Tayler cell, i.e. the gap width of the container, hence  the dynamo is  a small-scale dynamo. Magnetic energy slightly eceeds the  kinetic energy as it is typical for AMRI simulations for $\Pm\gsim 1$, see   \cite{RG18}. Also the numbers of the total torques $\nuT/\nu =$O(100)  are very similar in the simulation of the dynamo and the azimuthal magnetorotational instability


In a recent paper  Tayler-Spruit dynamo simulations are prsented in a spherical shell filled with a fluid of $\Pm=1$ \citep{Gissinger24}. The differential rotation is due to the rotation differences $\Delta \Om$ of inner and outer spheres, where $\Delta \Om/\Om = 3\ \E\ \Rey$ with fixed Ekman number $\E=$O($10^{-5}$). The Reynolds number $\Rey$ is defined with $\Delta \Om$ rather then the more convential value  $\Om$. The fluid is stably density-stratified where the Rayleigh number $\Ra$ directs the strength of the temperature gradient. The top panel of Fig. 3  of their paper, which presents simulations for $\Ra=10^7$ and with the weakest stratification, may exhibit a dynamo solution with toroidal and poloidal fields of rather small scales. It is a dipolar solution with the azimuthal field component  $B_\phi$ only  slightly exceeding the radial field $B_r$ very similar to the result of numerical simulations of the (nonaxisymmetric modes) of the pinch  instability without rotation \citep{RS11}. If a dynamo exists with differential rotation then the ratio of the azimuthal  and the radial field axisymmetric component should be basically larger than unity (see below).

There are also numerical simulations about the Tayler instability in rotating spheres with positive shear (``superrotation''). For Ekman number $\E=10^{-5}$ and magnetic Prandtl number $\Pm=1$ dynamo solutions have been found close to the inner tangent cylinder which are clearly small-scaled \citep{B23}. The pinch-type instability, however, is strongly suppressed by superrotation so that  only fields with small magnetic Mach number $\Mm<1$ become unstable under the conditions of Taylor-Couette flows. A subcritical excitation of the nonaxisymmetric disturbances  does only  exist or $\Pm\neq 1$  \citep{RS16}. Consequently, that the mentioned simulations provide an instability for superrotating fluids with $\Mm\simeq 100$ cannot be explained by consideration of simple Taylor-Couette flows.

\section{Basic equations}\label{stress}
The geometrical quantities $R$, $z$  and the wave numbers $k$ are normalized with 
\begin{equation}
R_0=\sqrt{(R_{\rm out} - R_{\rm in})R_{\rm in}},
\label{2.4}
\end{equation}
the frequencies with $\Omin$ and the magnetic fieds with 
$B_{\rm max}$. The unit of velocities is $\Omin R_0$. Then,
\beg
\Om(R)=a+{b \over R^2}
\label{e1}
\ende
with
\beg
a= {\mu-{\rin}^2 \over 1-{\rin}^2}\ \ \ \ \ \ 
 b={ \rin \over 1- \rin}
{1-\mu \over 1-{\rin}^2}
\label{e2}
\ende
with the normalized radius of the inner cylinder $\rin=\Rin/\Rout$ and the normalized outer rotation rate $\mu=\Omout/\Omin$. We shall only consider containers with $\rin=0.5$. Rigid rotation is described by $\mu=1$ while quasi-Keplerian rotation belongs to $\mu=\rin^{1.5}=0.35$.

 The radial profile of the azimuthal magnetic field may be the superposition  of two contributions.
The first  term -- called as the background field -- is formed by the solution   of the stationary induction equation without flow while the last term
formally describes the appearance of a magnetic ring with its maximum in the center of the TC-flow which is not a
formal solutions of any equation as also done in  \cite{KS14}):
\beg
B_\phi=a_B R+\frac{b_B}{R} + (B_{\rm max}-1) \sin^p({N\pi\frac{R\sqrt{(1-\rin)\rin}-\rin}{1-\rin}})
\label{e3}
\ende
with
\beg
a_B=\frac{1}{R_{\rm in}}  \frac{\rin(\mu_B-\rin)}{1-\rin^2},    \  \ \ \ \ \ \ \ \ \ b_B=R_{\rm in}\frac{1-\mu_B \rin} {1-\rin^2}
\label{e4}
\ende
with $B_\phi(\Rout)=\mu_B B_\phi(\Rin)$. The parameter $N$ marks the number of magnetic maxima  between $\rin$ and 1. $N=0$ generates the quasi-uniform background field without ring(s), 
$N=1$ generates one ring in the gap center while $N=2$ generates two rings of equal sign for even $p$ and of opposite sign for odd $p$. The axial  electric current which produces the quasi-uniform  background field with $B_{\rm max}=1$ and $\mu_B=1$ is positive-definite n the full gap and after (\ref{Tayler}) this field is unstable everywhere.  This is not basically true, however, for a structure with a distinct ring. Such fields are locally stable in the outer half of the container but they are not -- as we shall see --   in the global calculations.

We imagine these azimuthal magnetic fields as due to axial electric current systems which could be calculated but their knowledge is not  important. We shall consider below only two actual radial profiles. The first one forms an azimuthal field with a clear maximum in the gap center ($N=1$)  while the other one has maximum and minimum of the same strength ($N=2$). We shall also ask whether the fluctuations due to the Tayler instability of (\ref{e3}) are able to form an $\alpha$ effect  which may produce via an $\alpha\Om$ dynamo model  an axisymmetric azimuthal field similar to these forms. The answer which we shall give in the present paper is No. The dynamo-induced azimuthal field  always has much smaller axial scale than the original azimuthal field whose instability produces the radial profile of the $\alpha$ effect used. Additionally, we shall find that on the basis of the Tayler instability any  $\alpha\Om$ dynamo starting with one ring finally yields fields with two rings in radius.

By definition it is $B_\phi(\Rin)=1$ as the last term in Eq. (\ref{e3}) disappears. The outer value of the field follows from $B_\phi(\Rout)=\mu_B$.
For $N=0$ the ring disappears while its amplitude for $\h>1$  simply 
approximates $\h$. The exponent $p$ only manipulates the steepness of the profile. Varyation of  the factor $N$ in (\ref{e3})  yields an azimuthal  profile with $N$ symmetrical 
maxima (for even $p$). The question is how the  form and the amplitudes of the magnetic rings  influence 
the Taylor instability with respect to the critical magnetic amplitudes and the geometry of the cells.

The disturbances $\vec{u}$,  $\vec{b}$ and $p$ are developed into normal modes, 
\beg
[\vec{u},\vec{b},p]=[\vec{u}(R),\vec{b}(R),p(R)] {\rm exp}({{\rm i}(\omega t+kz+ m\phi)}).
\label{fluc}
\ende
 Here $k$ is the axial wave number of the perturbation, $m$ its azimuthal wave number and $\omega$ the complex frequency. The equations can be  taken from \cite{RS07}
with the Reynolds number 
\beg 
{\Rey}=\frac{R_0^2\Omin}{ \nu  }
\label{g11}
\ende
and the Hartmann number
\beg
{\Ha}=\frac{R_0 B_{\rm max}}{\sqrt{\mu_0\rho\nu\eta}}
\label{g12}
\ende
  with
 $B_{\rm max}$ as the maximum of the radial profile (\ref{e3}) which in a good approximation  equals $\h$ for $\h>1$. A related definition is $\Ha_0=\Ha(\Rey=0)$ which concerns the Hartmann number for resting fluids. We know from previous papers that for one and the same azimuthal field profile the resulting $\Ha_0$ does not depend on the value of the  magnetic Prandtl number. On the other hand, it is important to know the magnetic Mach number of rotation
 \beg
{\Mm}=\frac{\sqrt{\Pm}\ \Rey}{\Ha},
\label{g12}
\ende
which for almost all astropysical applications of the theory basivally exceeds unity. Only for magnetars the Mach number is smaller than unity.

The  boundary conditions of the flow fields are  the rigid ones, i.e. $u_R=u_{\phi}=u_z=0$ at
$R_{\rm in}=( \rin/ (1- \rin))^{1/2}$ and 
$R=R_{\rm out}=( \rin (1 - \rin))^{-1/2}$. For perfect-conducting walls the magnetic components must
fulfill the well-known conditions $b_R={ \d b_\phi/\d R}+ {b_\phi}/{ R}=0$. The azimuthal wave number is always fixed in this paper to $m=1$.



\section{Rigid rotation}

\begin{figure}
\vbox{
\includegraphics[width=0.8\columnwidth]{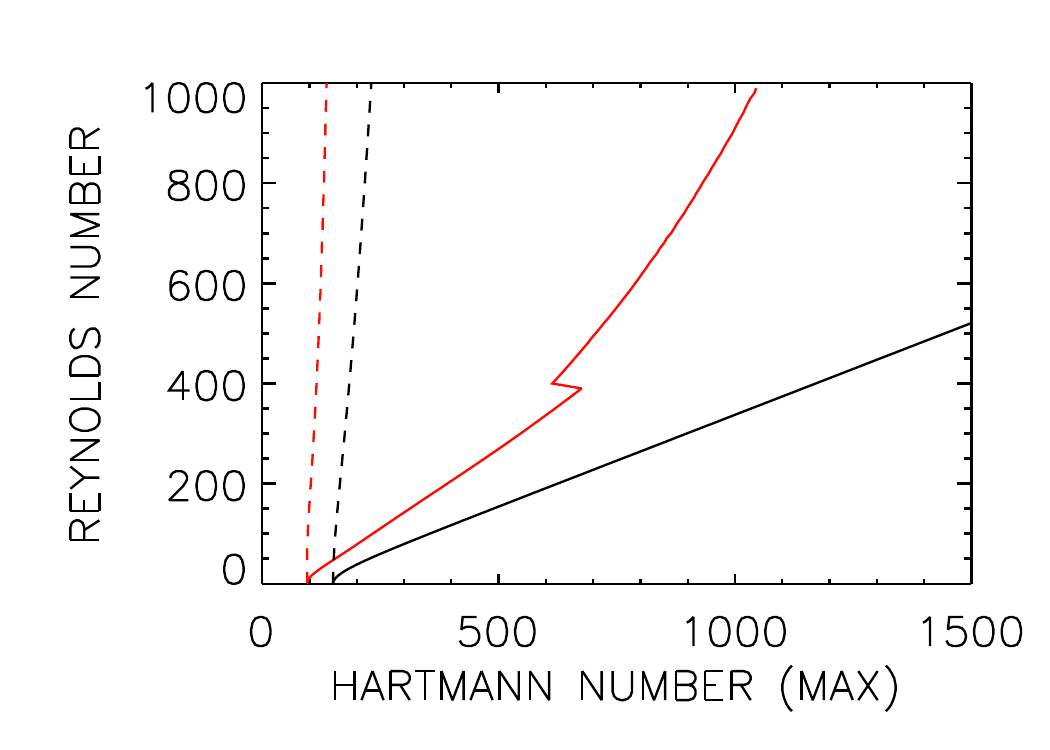}
}
\caption{The critical Reynolds numbers for marginal stability for two different radial profiles of the background field  with rigid rotation. The fields left from the lines are stable while they are unstable right from the lines.  It is $N=1$, $\h=1$ (black) and $\h=9$ (red, $p=3$). 
 $\Pm=1$ (solid),  $\Pm=0.001$ (dashed). $\rin=0.5$, $m=1$, $\mu_B=1$.}
\label{fig1a}
\end{figure}

In Fig. \ref{fig1a} for a given Hartmann number the black and red lines  for the various amplitudes of the ring define the minimal 
Reynolds numbers for which the magnetic field becomes stable. We start with a rigidly rotating container and $p=3$.  Figure \ref{fig1a}  gives the results for medium ($\Pm=1$, solid lines) and small ($\Pm=0.001$, dashed lines) magnetic Prandtl number.
The black curves represent the marginal  values for the smooth field without extra rings, they are already known from previous studies, \citep{RG18}. We  f note that   curves of the same colour for $\Rey=0$ have one and the same Hartmann numbers $\Ha_0\simeq 150$ independent of $\Pm$. Next the stabilizing influence of rigid global rotation is visible. However, the red curves  
for  strong magnetic rings are located  to the left of the black curves indicating  {\em destabilization} of the  field compared with the behaviour of quasi-uniform fields \citep{EB08}. We note that the red curves are much steeper for $\Pm=0.001$ than for $\Pm=1$, so that the destabilisation is enhanced for small magnetic Prandtl numbers. Hence, ring-like fields  and/or small $\Pm$ are always  more unstable than fields with uniform radial profile.  Additionally, the  curves of different colour are much more close together for small magnetic Prandtl numbers. Azimuthal magnetic fields - resting or rigidly rotating -  become unstable against nonaxisymmetric disturbances already for   lower magnetic field amplitudes if they exist in form of a magnetic ring rather than they are uniform in radius. This is  true for $\Pm\lsim 1$.  As the solid black and red lines demonstrate  for $\Pm=1$ the influence of the form of the magnetic background field for the rotational suppression of the Tayler instability is strong.  Typically, for equal Hartmann number a magnetic field of the ring-type is   unstable up to much higher Reynolds numbers than the smooth background fields are.
 We shall see in the following that the inclusion of rotational shear will considerably change these rules  we found for rigid rotation.
\section{Quasi-Keplerian rotation}
We apply now  differential rotation. We shall only consider the action of quasi-Keplerian rotation which for the model with $\rin=0.5$ is described by $\mu=0.35$. 

\begin{figure}
\vbox{
\includegraphics[width=0.8\columnwidth]{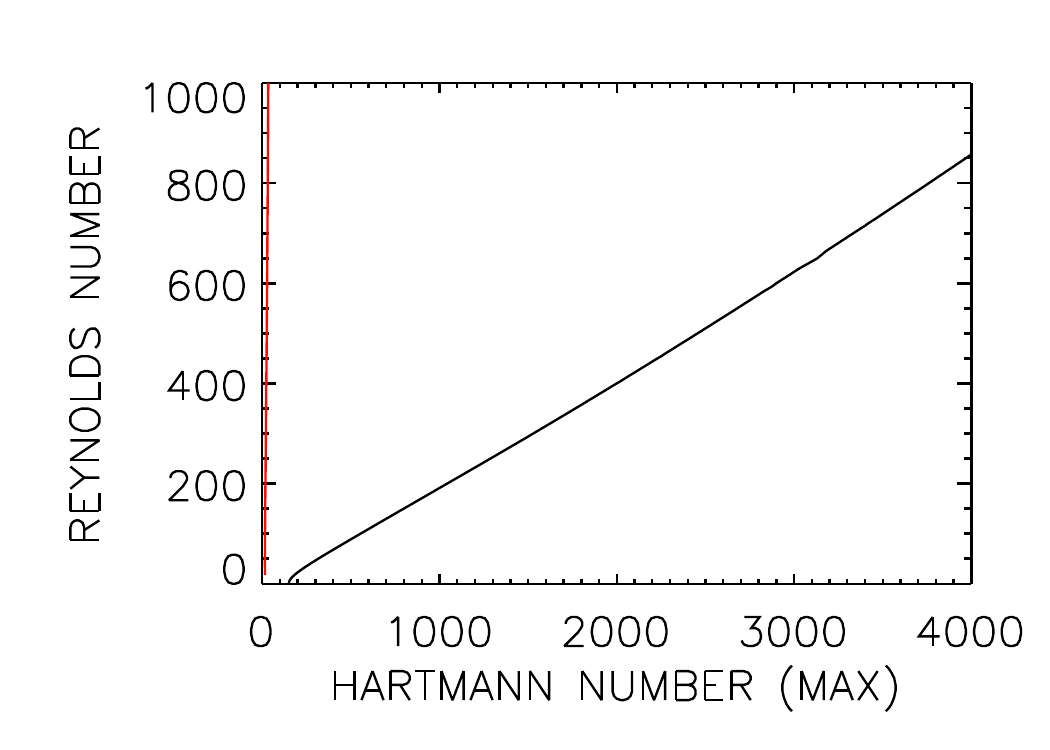}
\includegraphics[width=0.8\columnwidth]{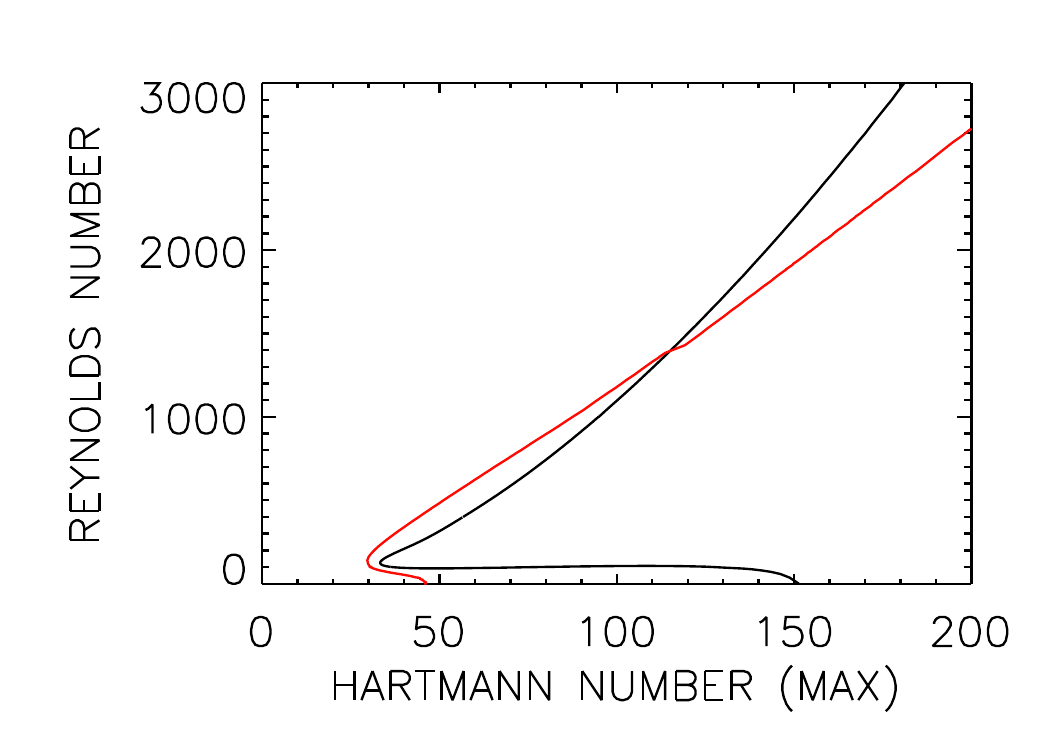}
\includegraphics[width=0.8\columnwidth]{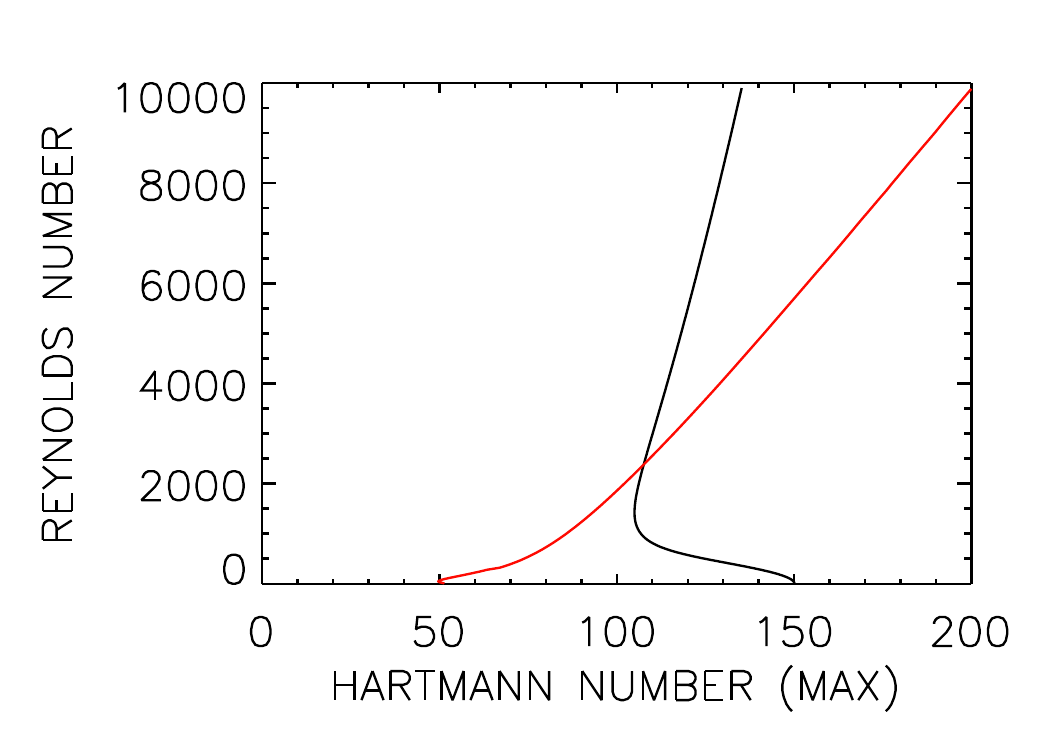}}
\caption{Similar to Fig. \ref{fig1a} but for  quasi-Keplerian  rotation  ($\mu=0.35$). Magnetic Prandtl number 
   $\Pm=10$ (top),  
$\Pm=1$ (middle) and $\Pm=0.001$ (bottom). It is $N=1$, $\h=1$ (black) and $\h=9$ (red), $m=1$,  $\mu_B=1$. 
}
\label{fig3a}
\end{figure}

Figures \ref{fig3a} demonstrate the influence of the shear flow onto the various field profiles with one maximum ($N=1$) for the three  magnetic Prandtl numbers $\Pm=10$, $\Pm=1$ and $\Pm=0.001$.  We  again note that for all panels the $\Ha_0$ of  the  black (quasi-uniform fields) and red (one central magnetic ring) lines do not depend on the magnetic Prandtl number. It also obvious that the lines of marginal stability under the influence of differential rotation are much steeper than they are for rigid rotation.  A shearing flow, therefore,  strongly destabilizes the magnetic field compared  to  the stabilizing influence of rigid rotation.

As the top panel  of Fig. \ref{fig3a} demonstrates the  instability of a quasi-uniform azimuthal background field  is {\em strongly} suppressed by the shear flow for large magnetic Prandtl numbers. The suppression is hampered, however, for ring-type fields with a clear maximum as the red line lies left of the black line.  Such fields are much more unstable against nonaxisymmetric perturbations  than the smooth fields, or - with other words -  the global rotation stabilizes the smooth fields much more effectively than the ring-type fields. More strictly speaking, the rotational stabilization of the nonaxisymmetric magnetic perturbations  is very weak for ring-type background fields so that such fields are highly unstable for not too large Reynolds numbers and  large $\Pm$. One finds that the stability map for large $\Pm$  {\em strongly} depends on the radial profile of the background field. For large magnetic Prandtl numbers ring-type toroidal magnetic fields should thus not survive the Tayler instability for $\Ha\simeq \Ha_0$. Models with large $\Pm$ must  thus be excluded from the dynamo-discussions of the present paper.

Generally, for $\Pm\lsim 1$ the red lines are located at the right side of the black lines, i.e. the rotation stabilizes toroidal fields of ring-type. The middle panel of Fig. \ref{fig3a} shows that the red line and the black line almost match for medium Reynolds numbers. As tests showed this remains also true  for all  $B_{\rm max}>5$. Under the influence of quasi-Keplerian rotation the ring-type field and the quasi-uniform field for the same magnetic  amplitude perform a very similar instability behaviour, i.e. the influences of the two steep gradients almost cancel each other. This is also true if the steepness $p$ of the radial magnetic profile is varied.  For sufficiently large $\h$ the critical Hartmann number does not change with the value of $\h$, also the influence of $p$ is weak. Hence, the form of the radial profile of the  field does here not play an important role in the instability map.  As the plot also shows this is not true   for resting models or models of low Reynolds numbers. One can find with Figs. \ref{fig1a} and \ref{fig3a}  the $\Ha_0$ of the red lines always smaller than the $\Ha_0$ of the black lines. For slow rotation the magnetic rings act always destabilizing.

Also for   smaller magnetic Prandtl number ($\Pm=0.001$, bottom panel) for medium and large Reynolds numbers the red line is  located at the right-hand side of the black line. The ring-type fields are thus more stable than the quasi-uniform field. For small magnetic Prandtl number and differential rotation we find the Tayler instability of magnetic rings  easily stabilized by  rotation. Also here the radial profile of the magnetic background field is  important for the lines of marginal stability. We find that only for $\Pm=1$  the form of the background field is not relevant for the rotational quenching of the Tayler instability. Models with small $\Pm$ are thus  excluded from the discussions in the present paper, too. For Tayler instability calculations the magnetic Prandtl number plays an important role.


\begin{figure}
\hbox{
\includegraphics[width=1.00\columnwidth]{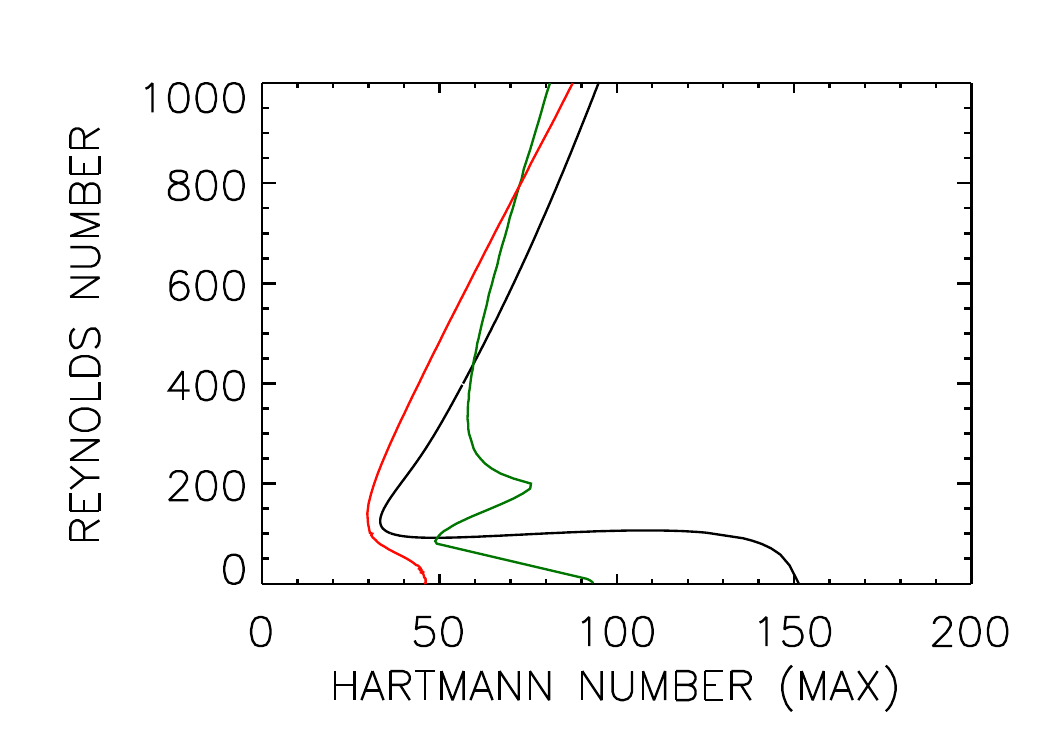}
}
\caption{
Summary of the stability maps for $\Pm=1$.   No ring (black line, $k=3.5)$), one ring (red line, $k=1.8$), two antiparallel rings (green line, $k=2.8$). Wave numbers taken at $\Rey=1000$.  }
\label{fig10a}
\end{figure}

 It is known that the rotation basically stabilizes the nonaxisymmetric Tayler instability. Figure \ref{fig10a} gives a summary of the lines of marginal instability for $\Pm=1$. The maximally possible Reynolds numbers for instability are plotted for given Hartmann numbers. The black line gives the well-known instability parameters  for $\mu_B=1$. For azimuthal fields with one ring (red) or two   rings (green) the curves  for finite Reynolds numbers are rather similar in opposition to the $\Ha_0$ for the non-rotating models. The rotation strongly reduces the differences of the excitation of perturbances of smooth or ring-like magnetic profiles. Drift frequencies are negative in all cases.  The magnetic Mach number at the upper margin of the plot sligthly exceeds 10.  With 1000 G and a density of $10^{-3}$g/cm$^3$ the Mach number of the solar NSSL exceeds 20 which \citep{Vasil24} is even  larger than the maximally possible Mach numbers shown in Fig. \ref{fig3a} for $\Pm=1$. It is thus questionable whether the solar NSSL can be considered as the site of a dynamo on the basis of  the  magnetic instability. Because of the higher density values the magnetic Mach number of the tachocline is about 200 which seems to be much too high for the appearence of Tayler instability of azimuthal field belts of (say) 1000 G.

\section{Alpha effect}\label{Alpha}
Consider the axisymmetric part of the  electromotive force ${\vec{\cal E}}=\langle \vec{u}\times \vec{b}\rangle$  which is due to the correlations of flow and field perturbations. The outline which we are following has already been developed by \cite{RS20}.
The line of neutral stability defines the weakest  possible rotation rates  ensuring stability of a given magnetic field. 

The averaging procedure concerns the time and the  coordinates $\phi$ and $z$. The mean-field  electromotive force may be developed in the  series
\beg
{\vec{\cal E}}= \alpha \vec{B}- \eta_{\rm T} \curl \vec{B}+....
\label{EMF}
\ende
with the $\alpha$ effect and the eddy diffusivity   $ \eta_{\rm T}$.  In cylinder geometry, the $\phi$-component
$
{\cal E}_\phi= \langle u_z b_R -  u_R b_z\rangle 
$
 can be written as  the axisymmetric part of expressions  such as $\langle u_z b_R\rangle\propto u_z^{\rm R} b_R^{\rm R}+u_z^{\rm I} b_R^{\rm I}$. 
It is thus evident that the total azimuthal  electromotive force due to the   instability of azimuthal fields  vanishes if both modes $m$ and $-m$ (which have the same eigenvalues and the same azimuthal drifts) are simultaneously excited with the same power. Only   by an extra parity braking (e.g. by an additional $z$-component of the magnetic background field) finite values of the $\alpha$ effect appear. Another possibility is to consider the isolated modes $m=1$ and $m=-1$   as the result of  a spontaneous parity braking or  by use of strictly formulated initial conditions. 
 If the initial conditions clearly  favor one mode then only   this one is excited. If the initial condition do not favor one of the two modes, the numerical noise will excite all the possible modes.  

Consider   the components 
\beg
{\cal E}_\phi= \langle u_z b_R -  u_R b_z\rangle 
\label{Ephi} 
\ende
and   
\beg
{\cal E}_z=\langle u_R b_\phi -  u_\phi b_R\rangle
\label{Ez}
\ende
 of the electromotive force. Only the ratio of both quantities is free of arbitrary factors and/or  normalizations. 
We write $\varepsilon_\alpha={\cal E}_\phi/{\cal E}_z$, i.e.
\beg
\varepsilon_\alpha=\frac{ \langle u_z b_R -  u_R b_z\rangle }{\langle u_R b_\phi -  u_\phi b_R\rangle}.
\label{epsilon}
\ende
With ${\cal E}_\phi=\alpha B_\phi$ and ${\cal E}_z=-\eta_{\rm T} \curl_z \vec{ B}$ one immediately finds 
$
{C^{\rm sim}_\alpha} =-\varepsilon_\alpha \ \delta
$
for the dimensionless number
\beg
C^{\rm sim}_\alpha= \frac{\alpha R_0}{\eta_{\rm T}},
\label{EMF3}
\ende
which in the dynamo theory  governs  the excitation of dynamo-generated magnetic fields. For the normalized current it is 
\beg
\delta= 1+ \frac{R}{B_\phi}\frac{\d B_\phi}{\d R}
\label{EMF4}
\ende
As an example it is $C^{\rm sim}_\alpha=-2 \varepsilon_\alpha $ for the pinch field with $\mu_B=2$. The radial profiles of 
the  $\delta$ for the azimuthal fields can simply be calculated. It it aproaches unity for the quasi-uniform field with $\mu_B=1$  while for the one-ring profile the maximal values fulfill $|\delta|<4$.

\begin{figure}\vbox{
\hbox{
\includegraphics[width=0.5\columnwidth]{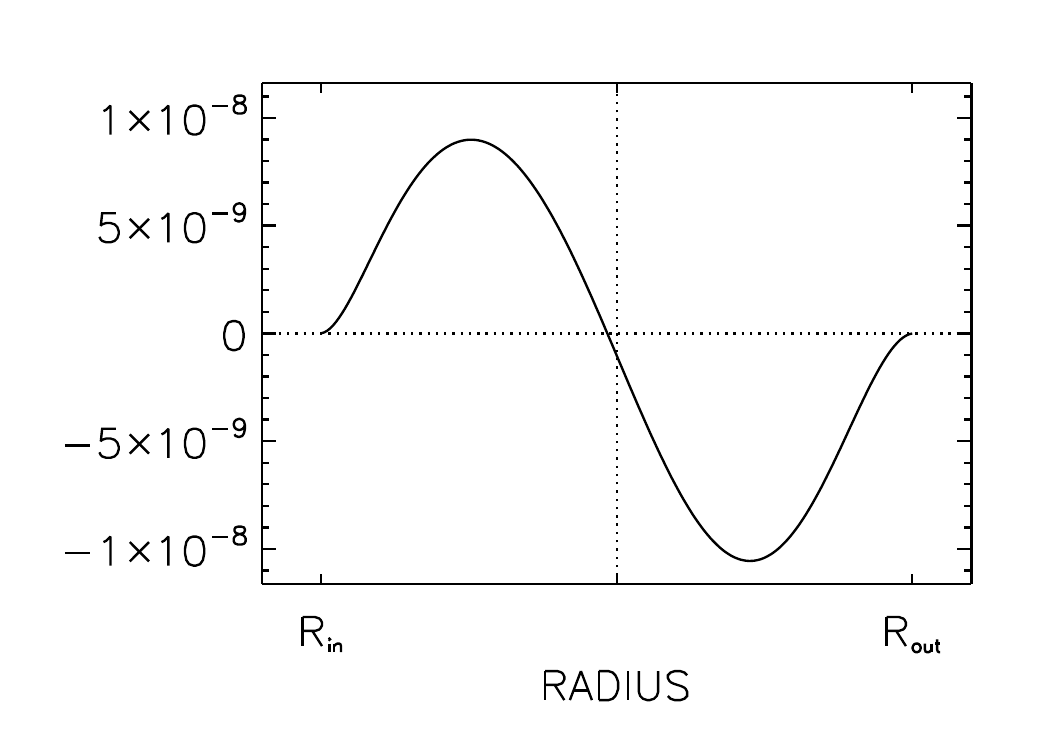}
\includegraphics[width=0.5\columnwidth]{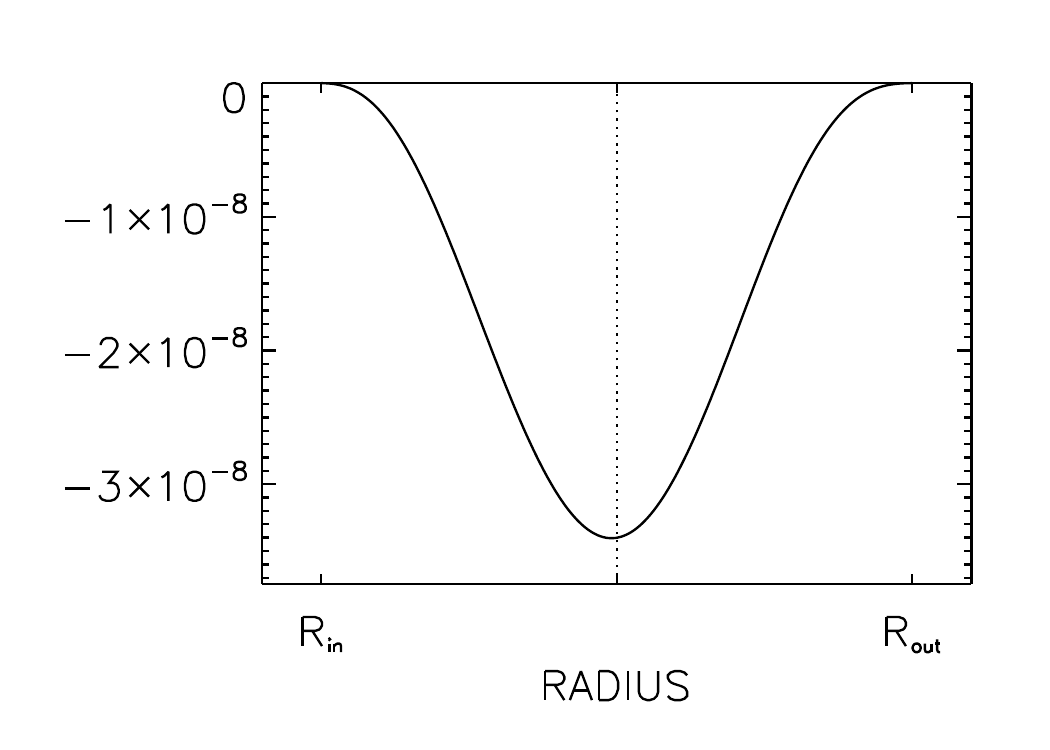}
}
}
\caption{Radial profiles of  ${\cal E}_\phi=\alpha B_\phi$ (left) and ${\cal E}_z=-\eta_{\rm T} \curl_z \vec{ B}$ (right) in arbitrary units for $\Rey=0$.   Quasi-uniform toroidal magnetic field with $\mu_B=1$, $k=4.4$, $\Ha=150$, $\Pm=1$.
 The $C_\alpha$ profile is  of the sine-type changing  sign in the gap center  and leads to the amplitude $C_\alpha\simeq 0.5$. The $\etaT$ is positive-definite.}
\label{fig6b}
\end{figure}

We start with the maximal simple model without rotation ($\Rey=0$) and for quasi-uniform magnetic field with $\mu_B=1$ (Fig. \ref{fig6b}, upper row).  The left panel  gives the radial profiles of (\ref{Ephi}), i.e.  ${\cal E}_\phi=\alpha B_\phi$  and the right panel provides (\ref{Ez}), i.e. ${\cal E}_z=-\eta_{\rm T} \curl_z \vec{ B}$  in arbitrary units. Only the ratio of both quantities has a physical meaning. Because of the simple form of the quasi-uniform magnetic field and the simple form of the positive quasi-uniform electric current the interpretation of the results is rather simple. The left-hand plot represents the $\alpha$ effect while the right plot represents the (negative) eddy diffusivity $\eta_{\rm T}$. Hence, even for nonrotating containers an $\alpha$ effect exists with two different signs along the radius and with a zero in the gap center. The  signs given in the plot only depend on our choice that $m=1$, they would be opposite for $m=-1$. The finite azimuthal electromotive force (i.e. the $\alpha$ effect) would thus vanish if both modes are simultaneously   excited. In the given example  they also vanish for each single mode if averaged over the radius. The latter follows from the symmetry of   $B_\phi$ with respect to the gap center. It would  not be true if the background field failed this symmetry as for example the background field  of  pinch-type with $\mu_B=2$ as studied by \cite{RS20}. Nevertheless, that the Tayler instability in cylindric geometry always leads to an $\alpha$ effect of both signs between the boundaries is a very robust result of the calculations with severe consequences for a possible dynamo process.  The helicity changes its sign  at a particular radius in opposition to the well-known helicity of rotating convection which changes its sign at a particular latitude, i.e., at the equator. Also by nonlinear calculations of  \cite{AR11} the different signs of the $\alpha$ effect along the distance from the axis appeared in spherical geometry.

The opposite is true for the resistivity $\etaT$. The right  panel of Fig. \ref{fig6b} demonstrates its positivity as the electric current is positive-definite for $N=0$ while it changes its sign in the middle of the gap for $N=1$  so that $\etaT>0$ follows with its  maximum close to  the center.   This result does not depend on the special choice of the azimuthal mode number $m$.


\begin{figure}
\vbox{
\hbox{
\includegraphics[width=0.5\columnwidth]{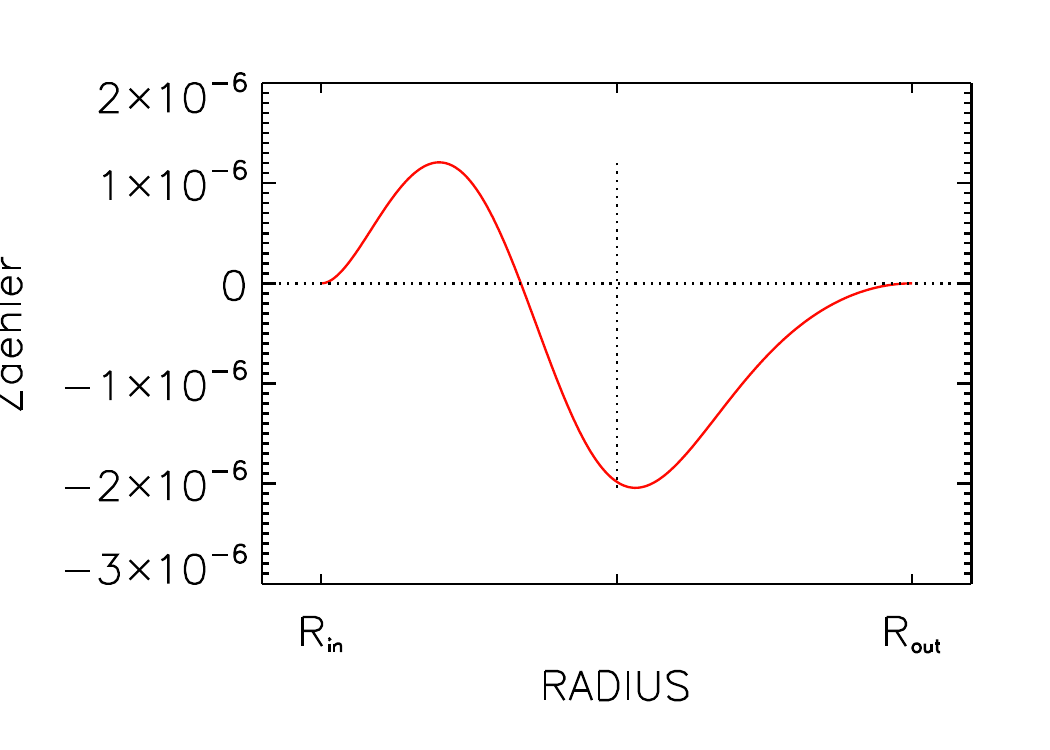}
\includegraphics[width=0.5\columnwidth]{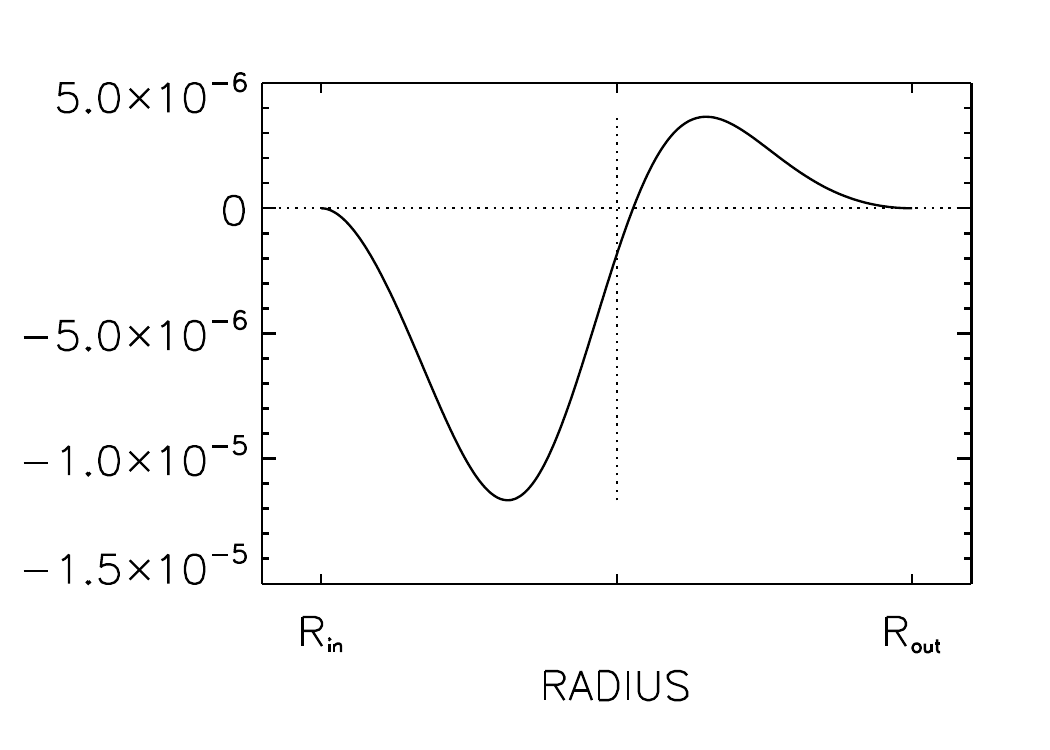}}
\hbox{
\includegraphics[width=0.5\columnwidth]{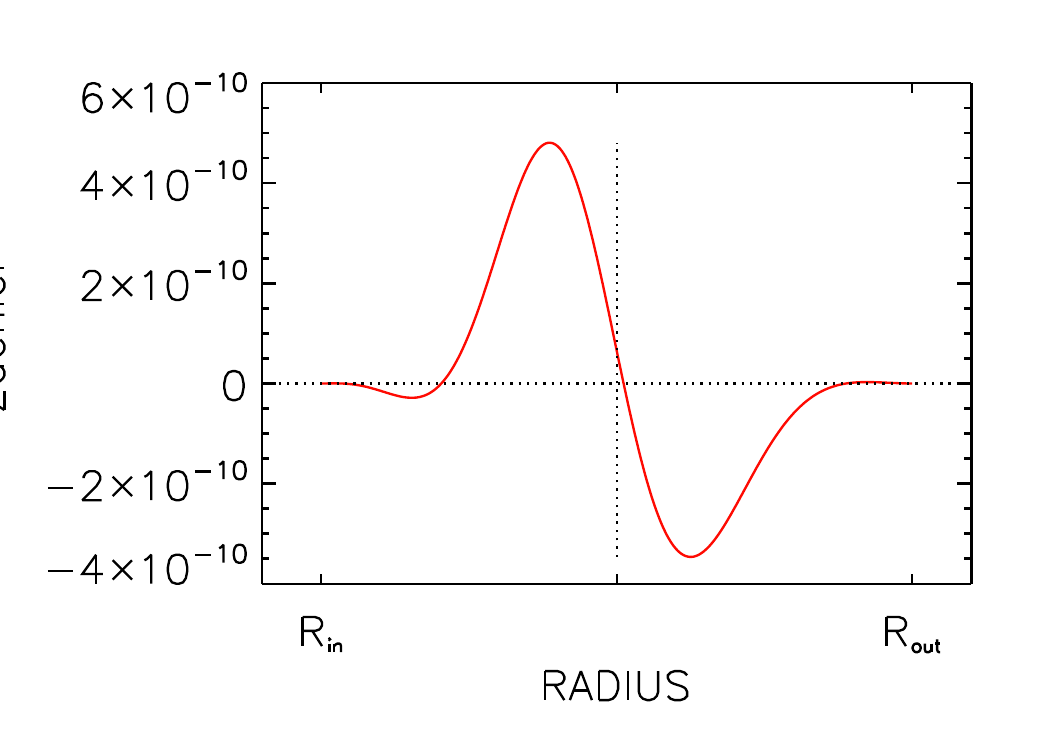}
\includegraphics[width=0.5\columnwidth]{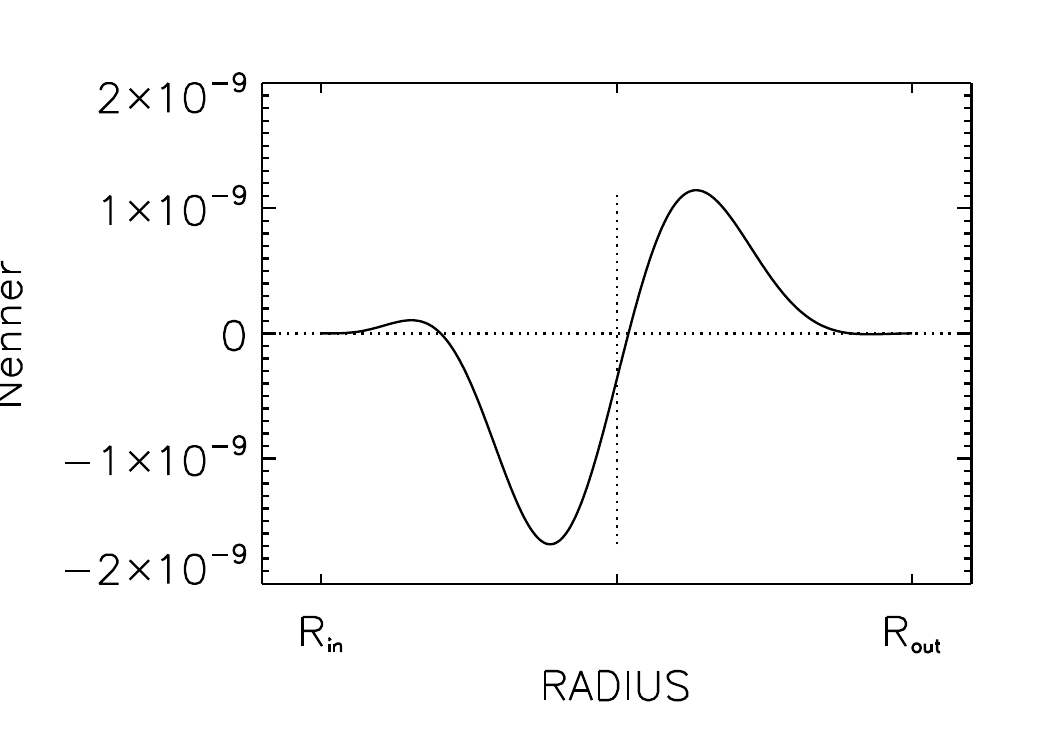}
}
\hbox{
\includegraphics[width=0.5\columnwidth]{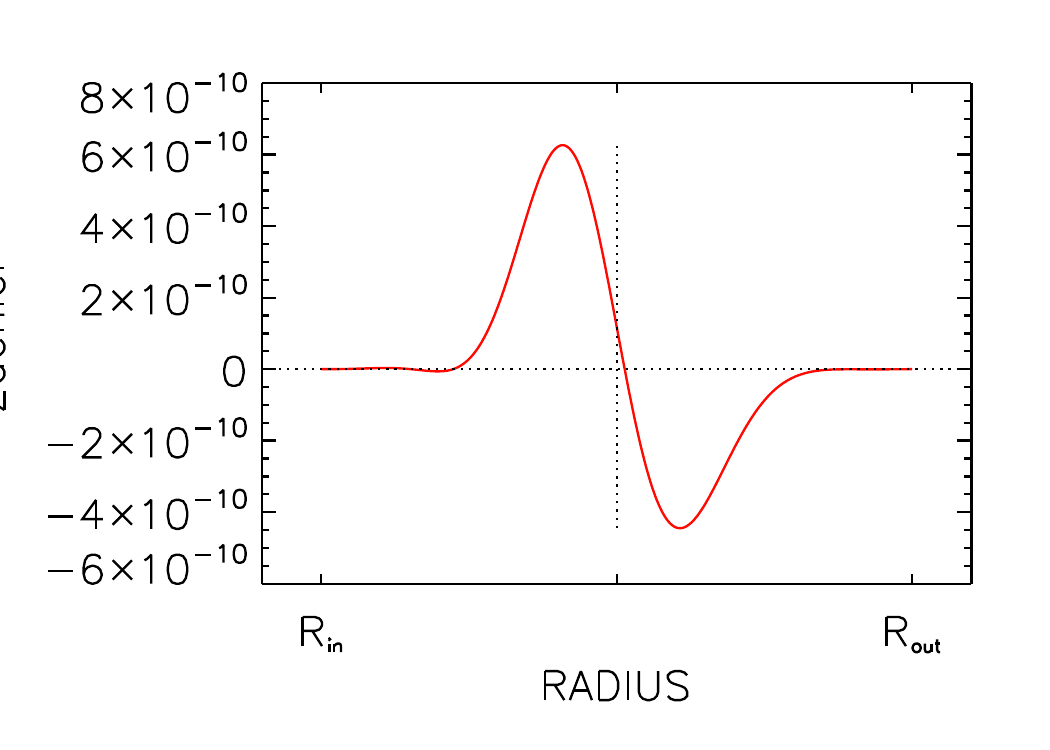}
\includegraphics[width=0.5\columnwidth]{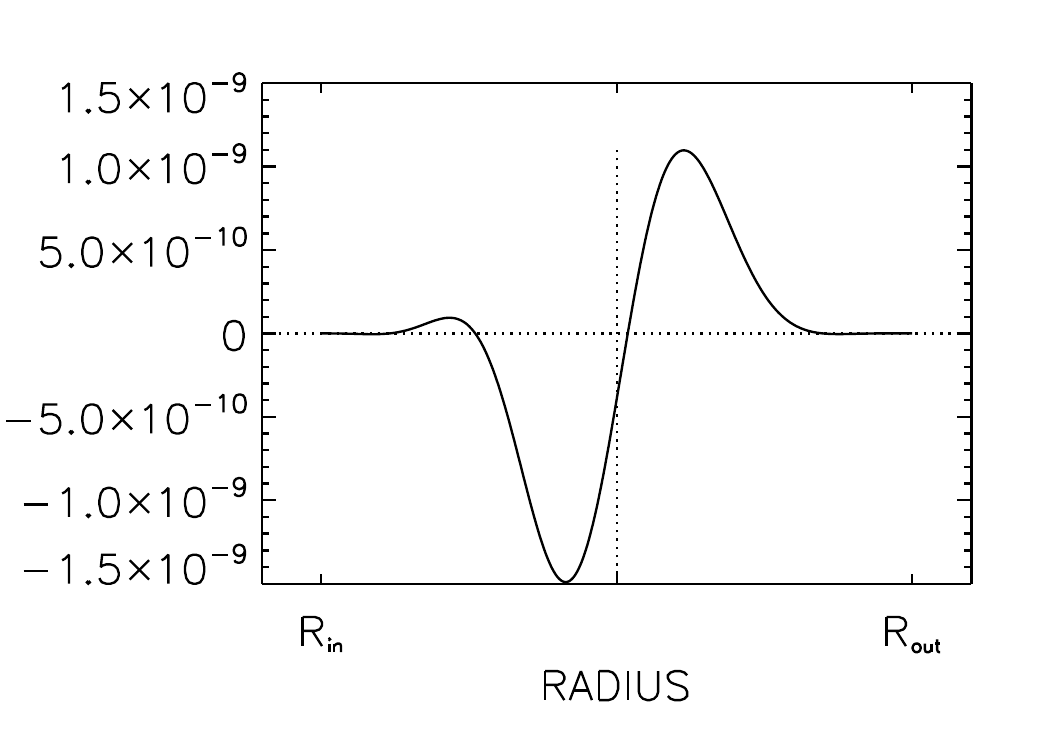}}
}
\caption{Similar to Fig. \ref{fig6b} but for the one-ring background field after (\ref{e3}) with  $N=1$,  $\h=8$ and for quasi-Keplerian rotation. Top: $\Rey=0$, middle:   $\Rey=1000$,  bottom:  $\Rey=2000$.   The $C_\alpha$ is always of the sine-type with the zero close to  the gap center.   $m=1$, $\Pm=1$. } 
\label{fig7a}
\end{figure}

Very similar correlations  for the electromotive force are provided by rotating models. We shall only consider the results for models with quasi-Keplerian rotation law which are located at the upper branch of the instability map (Fig. \ref{fig3a}) so that the maximally possible rotation rate is considered, i.e.  $\Mm\lsim 10$. For faster rotation the Tayler instability  disappears. The  magnetic background field is always positive  and has  no  ($N=0$) or only one ring ($N=1$) with $\Pm=1$.    The  panels of Fig. \ref{fig7a} provide the results  for the one-ring profile with $N=1$.  By the left column the azimuthal electromotive force (\ref{Ephi}) is presented with again  the same structure with two different signs  within the cylinders as we already obtained for $\Rey=0$.  As the applied magnetic field (\ref{e3}) is   positive-definite in both cases, also for rotating containers  the $C_\alpha$  in the gap  has alternating signs  with a zero just in the gap center.  This result also holds for other values of the Reynolds number.

The interpretation of the  axial component   ${\cal E}_z=-\eta_{\rm T} \curl_z \vec{ B}$ of the electromotive force bases on the fact that for the ring-type field the axial electric current is positive (negative) in the inner (outer) part of the container. As the signs in the plot on the right-hand side  of  the bottom row are opposite, the coefficient $\eta_{\rm T}$ in this relation must be positive. Again, the positivity of the dissipation quantity does neither depend on the rotation rate nor on the form of the background field.

The simulated normalized $\alpha$ effect  (\ref{EMF3}) in all cases proves  to be of order unity. For $\Pm=1$ one finds $C_\alpha\simeq 0.75$ and  $C_\alpha\simeq 0.80$ for $\Rey=1000$ and $\Rey=2000$. 
  The influences of the Reynolds number and the particular radial profile of the magnetic backgroud field are only weak. The  sinusoidal form of the $C_\alpha$ profile - changing sign close to the  gap center - is very robust. It  exists for nonrotating quasi-uniform magnetic fields but  also for a rotating container with a one-ring magnetic background field. For small $\Pm$ the only modification is that the profile becomes asymmetric because of the migration of the central zero in the direction of the inner cylinder.


 The normalized instability  frequencies $|\omega|/\Om$ always remain smaller than 0.5, what means that the turnover time of the perturbation is always shorter than the rotation time. 

 A general rule exists  for the  axial scales of the instability patterns. The red line of the middle panel of Fig. \ref{fig3a} yields  for $\Rey=1000$ the wave number  $k=1.5$. In this case the background field only possesses one maximum ring.  For fields with two rings of opposite polarity an axial wave number of  $k=2.8$ results. The axial scale is   reduced the more structured the background field is. The axial scales within  the induction mechanism become a cascade towards small scales which might be characteristic for a decay process.

\section{Mean-field dynamos}\label{Dyn}
We return to the dynamo models in cylindric containers with $\rin=0.5$ which  have been constructed in \cite{RS20}. From numerical reasons we are focussed here to the case of $\Ha\simeq 100$ (corresponding to 200 G within a TC-flow of sodium with 10 cm gap width) which after Fig. \ref{fig10a} leads to a maximally possible rotation with $\Rey\simeq 1000$. The corresponding  solutions are   derived of drifting $\alpha\Om$ dynamos with  different radial profiles of the $\alpha$ effect plotted in the left column of Fig. \ref{fig9a}. The  $\alpha$ effect with the zero in the gap center  is just the type which we found   as due to the  Tayler instability of an azimuthal field with no  or with only one ring. Because  the smallness of the $\alpha$ effect by Tayler instability the construction of $\alpha^2$ dynamos in cylindrical geometry is impossible \citep{RS20}. The  remaining $\alpha\Om$ dynamos need a steep rotation law so that always $|B_\phi| \gg |B_R$ is a characteristic property of any  mean-field dynamo on the basis of the Tayler instability (without density stratification). The almost equal amplitudes of toroidal field and poloidal field of the given solution in the upper line of Fig. 3 of \cite{Gissinger24} basically  contradict  this finding for Tayler-Spruits dynamos.

\begin{figure}
\includegraphics[width=1.00\columnwidth]{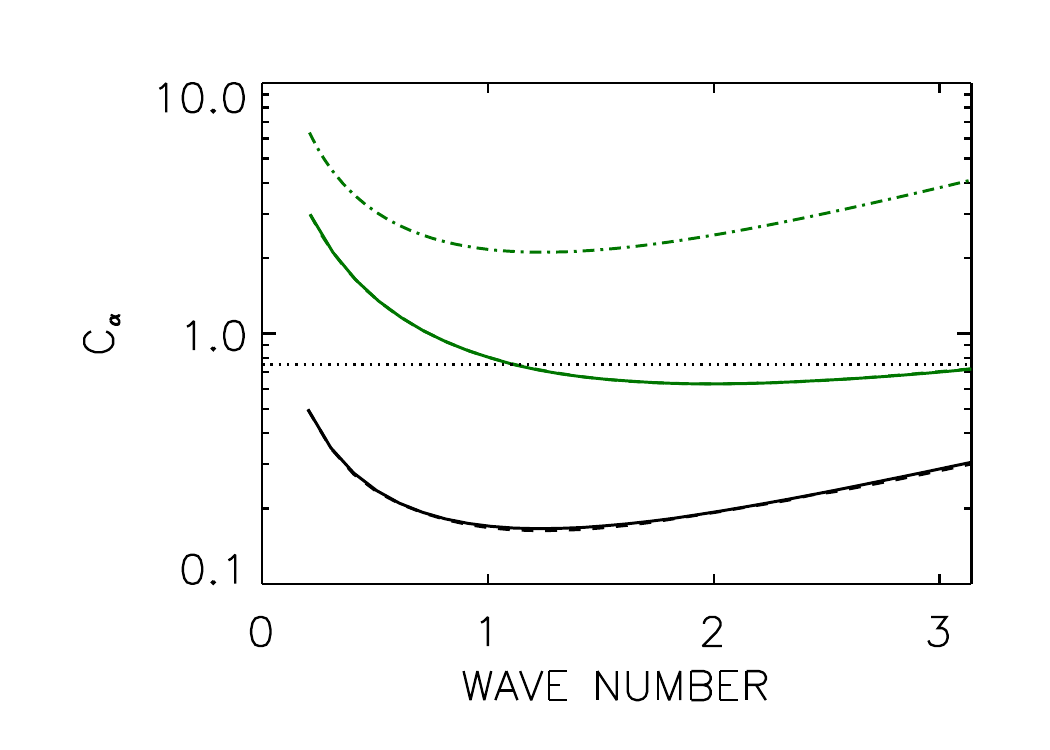}
\caption{Critical amplitudes of  $C_\alpha$ of a cylindric  $\alpha\Om$ dynamo with quasi-Keplerian rotation law for $C_\Om=1000$  vs. the vertical wave number $K$. Black:  artificial quasi-uniform  $\alpha$ effect, green: sin-type $\alpha$ effect with a central node. The  area between the black line and the green line contains the marginal $C_\alpha$ amplitudes for all radial profiles with one node independent of its location.  The dot-dashed line gives the eigenvalues for $\alpha$ effect vanishing in the inner half of the container. The horizontal dotted line represents the $C_\alpha=0.75$ typical for Tayler instability. The curves are nearly identical with the solutions after $\alpha \to -\alpha$  (dashed lines). Vacuum boundary conditions. }
\label{fig9}
\end{figure}
\begin{figure}
\includegraphics[width=1.00\columnwidth]{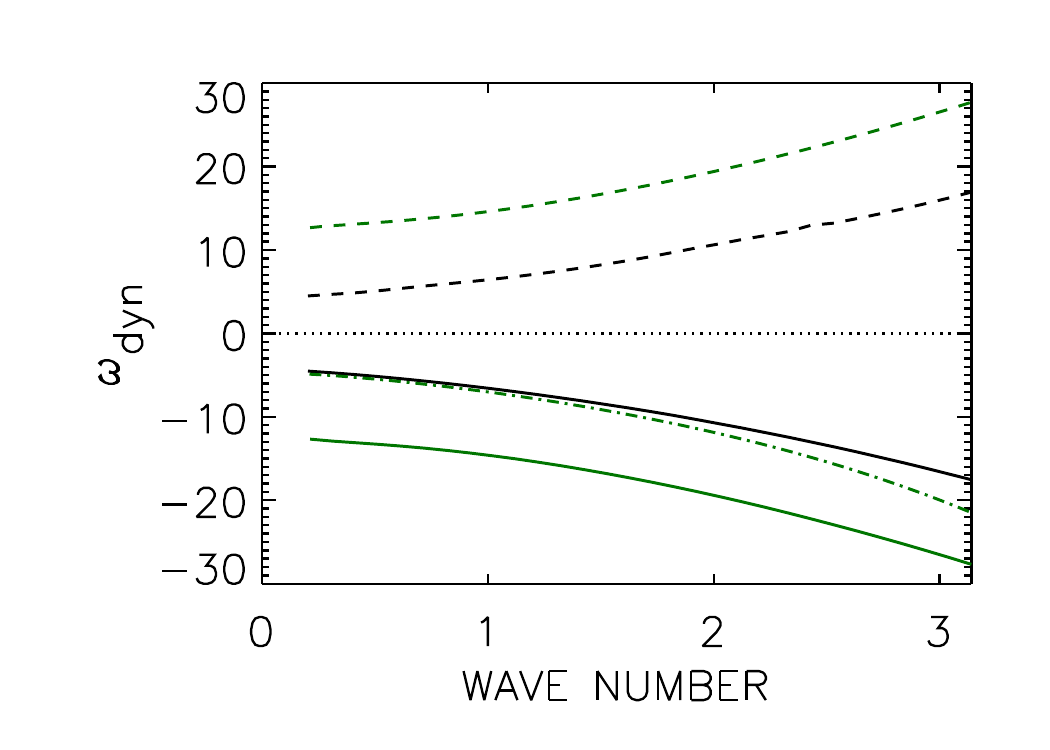}
\caption{Similar to Fig. \ref{fig9} but for the normalized drift frequency of the dynamos. The one-ring dynamo exhibits the fastest oscillations.}
\label{fig9d}
\end{figure} 

Figure \ref{fig9} gives the dependence of the critical dynamo number $C_\alpha$ on the axial wave number $K$ of the dynamo   for marginal excitation of an cylindric  $\alpha\Om$ dynamo with  $C_\Om=1000$ where as usual $C_\Om= \Omin R_0^2/\etaT$ so that 
$C_\Om\simeq \Pm \Rey\ \eta/\etaT$. The following considerations are often done for small  $\etaT\simeq \eta$. Dynamos with $K=0$ are not possible so that infinite values of $C_\alpha$ appear. The axial wave number $K$ is normalized with the gap width $D=\Rout-\Rin$ which for our model with $\rin=0.5$ is identical with the above normalization factor $R_0$, hence we can directly compare the wave numbers $k$ and $K$. This is not true for the frequencies, as the different normalizations provide for the ratio of the frequencies of the perturbation waves $\omega_{\rm T}$ and the cycle frequency $\omega_{\rm cyc}$ of the dynamo as
 \beg
\frac{\omega_{\rm T}}{\omega_{\rm cyc}} = \Pm \Rey \frac{\omega_{\rm dr}}{\omega_{\rm dyn}},
\label{omegas}
\ende
taken for $\etaT=\eta$. Here $\omega_{\rm dr}$ is the usual drift frequency of the instability normalized with the rotation rate and $\omega_{\rm dyn}$ is the dynamo frequency normalized with the diffusion frequency. For mean-field dynamo solutions the number (\ref{omegas}) must be large what is certainly fulfilled  for $\Pm=1$ and $\Rey=1000$ but it is certainly not fulfilled for $\Pm=0.001$. The existence  of such oscillating  dynamos for $\Pm\ll 1$ still remains  unclear. At $\Pm=0.001$ we needed the electromotive forces of Tayler instability at $\Rey=10^6$ in order to probe a dynamo with $C_\Om=1000$.

The critical amplitudes  $C_\alpha$ for different radial profiles of the $\alpha$ effect are given in Fig. \ref{fig9} as function of the wave number $K$. The black line describes the marginal dynamo excitation  of a dynamo for (artificial)  semi-positive quasi-uniform $\alpha$ effect with $\mu_B=1$. This profile provides the easiest excitation conditions, the minimal $C_\alpha$ belongs to $K\simeq 1.2$  what means periodic magnetic patterns on a distance of about $2\pi D$. If the $\alpha$ profile is of the sine-type as it is characteristic for the Tayler instability then - as the green line shows - the needed $C_\alpha$ are higher by one order of magnitude and  the  minimum of the dynamo number  moves to  $K\simeq 2.0$ what means periodic magnetic patterns on a distance of only $\pi D$. Compared to the black line for the  too simple $\alpha$ model the green line  leads to  higher dynamo numbers with minima at smaller scales. Only  dynamo models at the green line with $K\gsim 1.2$ are supercritical.

We note that the instability pattern of the azimuthal field with one ring also possesses a wave number of $k=1.8$, see Fig. \ref{fig10a}. Therefore, the mean-field approach  typically  leads to  the excitation of  small-scale dynamos with 
\beg 
K\simeq k,
\label{Kk}
\ende
 as has also numerically been demonstrated with a Taylor-Couette flow  
\cite{GH17} and  with a spherical model with superrotation \citep{B23}. The plot also shows that self-excited dynamos with larger scales, $K\ll 1.2$, needed $C_\alpha\gg 1$ which, however,  do not exist. After our results a  large-scale mean-field dynamo is  impossible on the basis of the Tayler instability as the helicity is not large enough. The dynamo solutions which we found are only operating at nearly  the same scale as the Tayler instability itself.  We note that as shown by  Fig \ref{fig3a} (middle)  the dynamo calculations are made for the highest possible rotation rate ($C_\Om=1000$) beyond which the  instability disappears. 

 For the mode with $m=-1$ the $\alpha$ effect  has  opposite signs ($\alpha \to -\alpha$) but the $\alpha\Om$ dynamo is not invariant against that 
 transformation \cite{D93}. For  $\alpha\Om$ dynamos with opposite signs of $\alpha$, however,  the  curves of marginal excitation are here almost identical with the original lines (Fig. \ref{fig9}).   The drift frequencies, however, for both solutions have opposite signs (Fig. \ref{fig9d}). Hence, if both modes exist with the same power the total $\alpha$ effect is zero and a dynamo cannot not exist. We note that the green lines in Fig. \ref{fig9d} for an $\alpha$ effect of the sine-type provide much higher drift frequencies than the black lines for  quasi-homogeneous $\alpha$ profiles. This result may imply that the cycle times of oscillating Tayler-Spruit dynamos are very short compared to the diffusion time.

\begin{figure*}
\hskip0.25cm
\vbox{
\hbox{
\includegraphics[width=0.65\columnwidth]{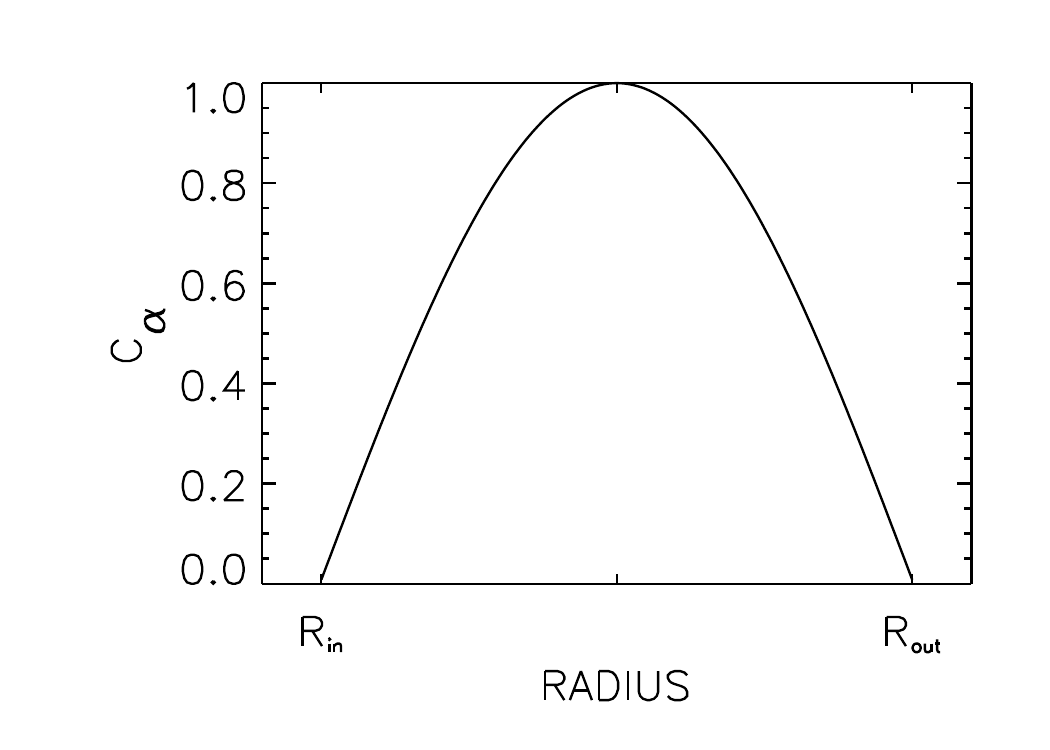}
\includegraphics[width=0.65\columnwidth]{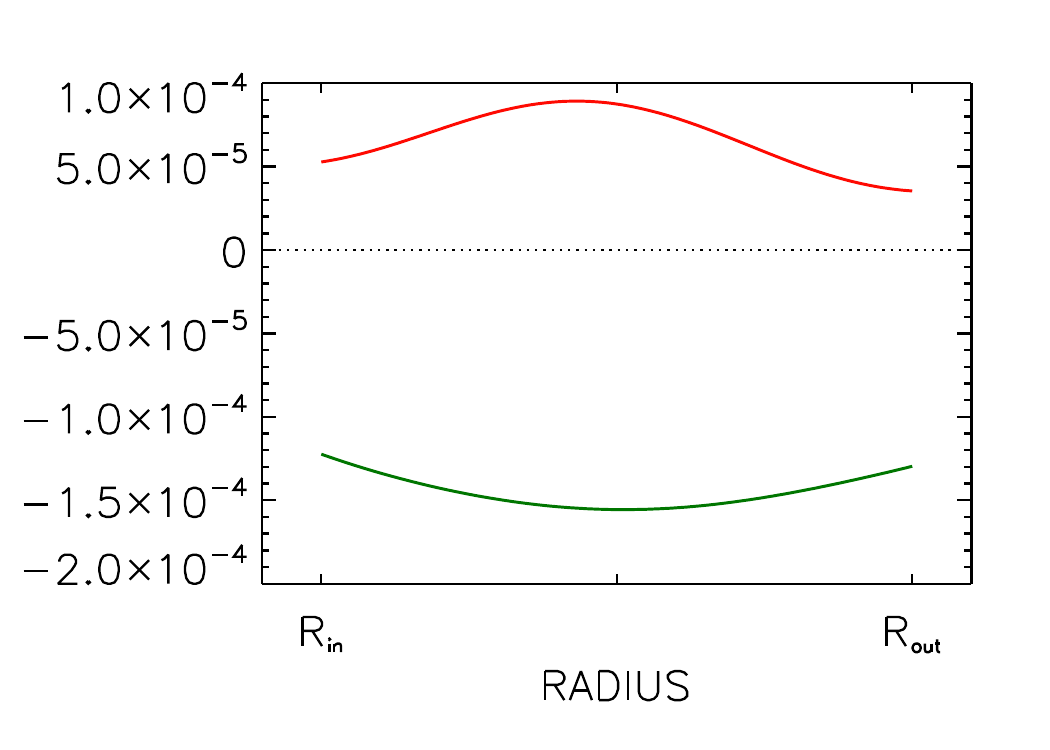}
\includegraphics[width=0.65\columnwidth]{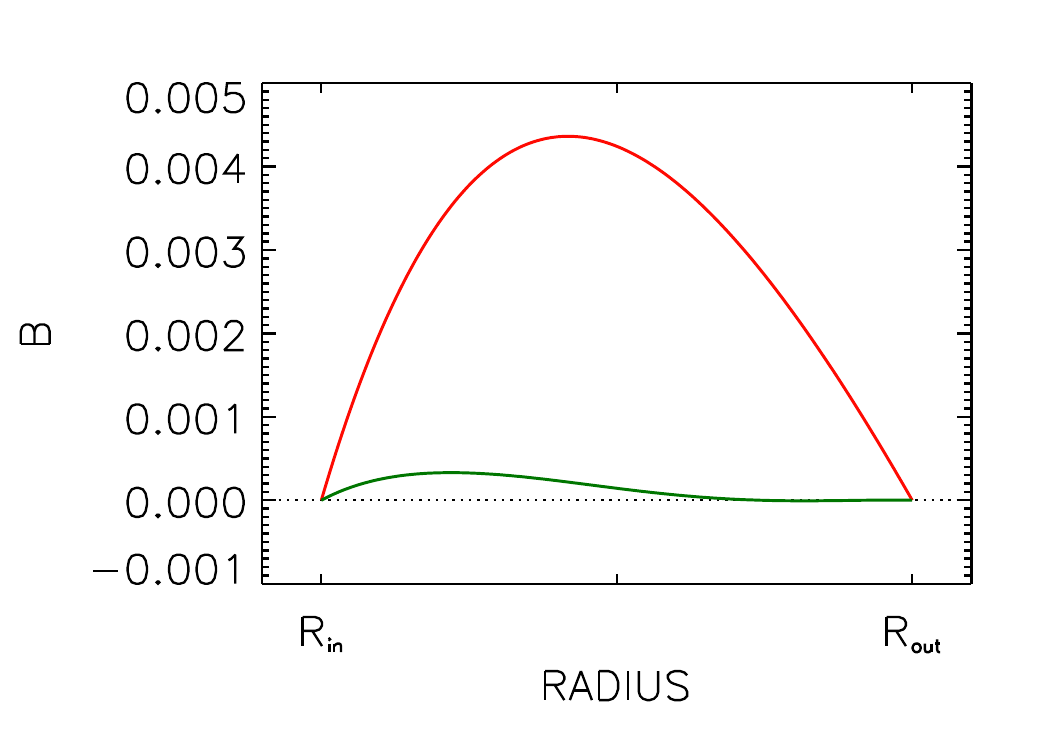}
}
\hbox{
\includegraphics[width=0.65\columnwidth]{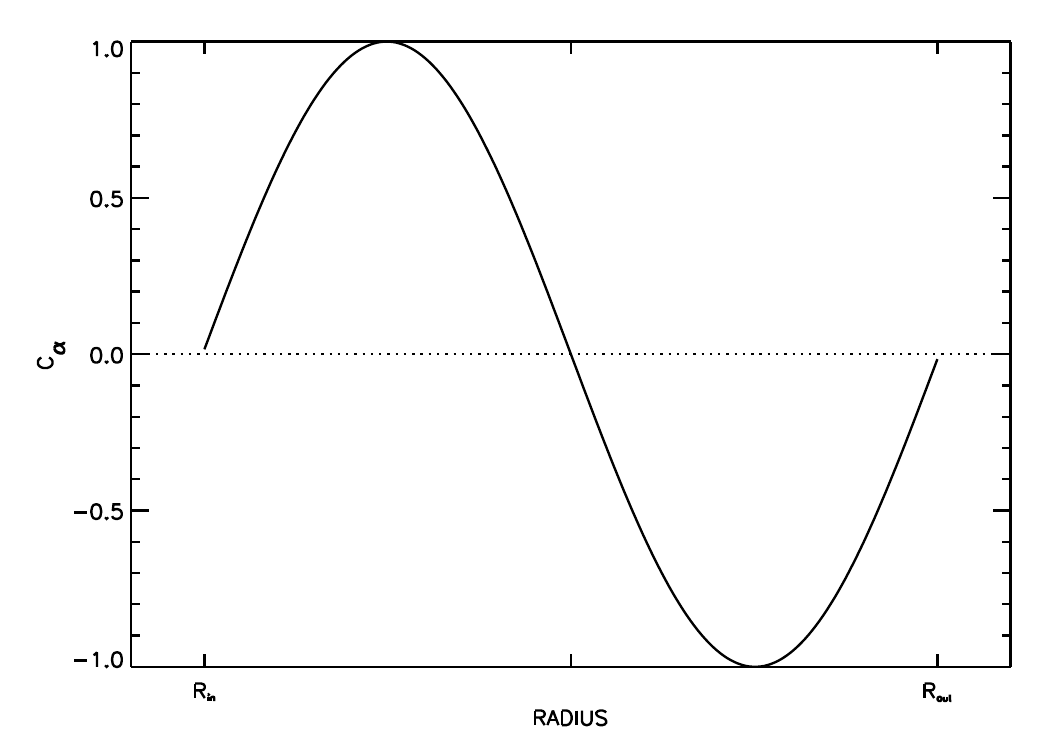}
\includegraphics[width=0.65\columnwidth]{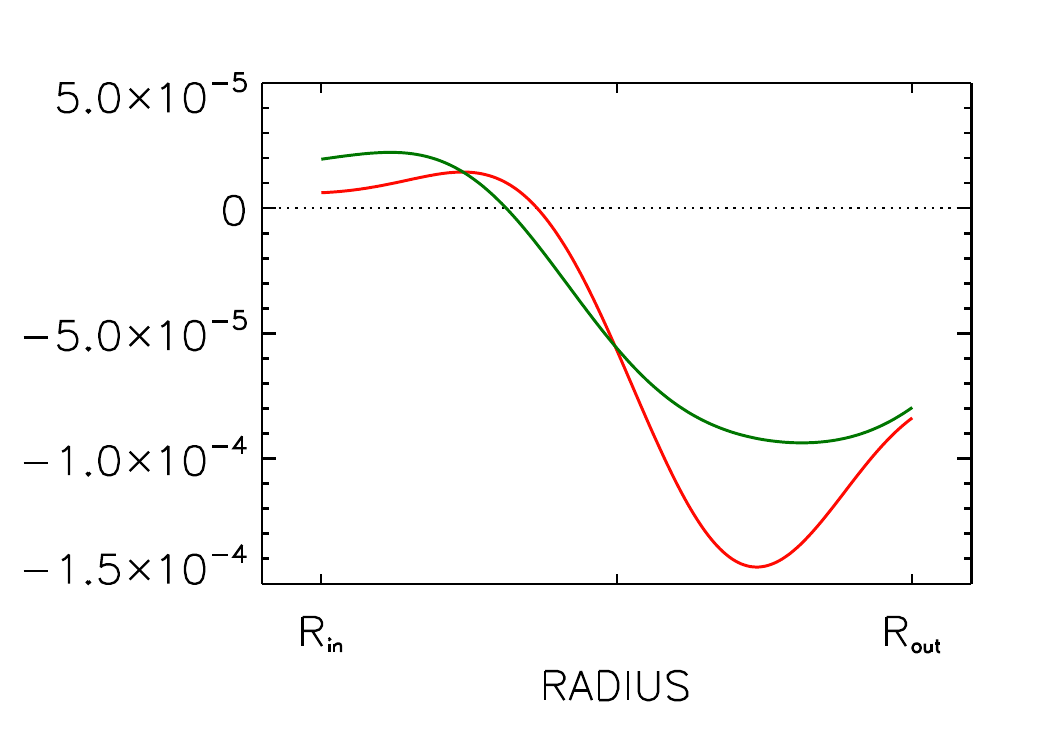}
\includegraphics[width=0.65\columnwidth]{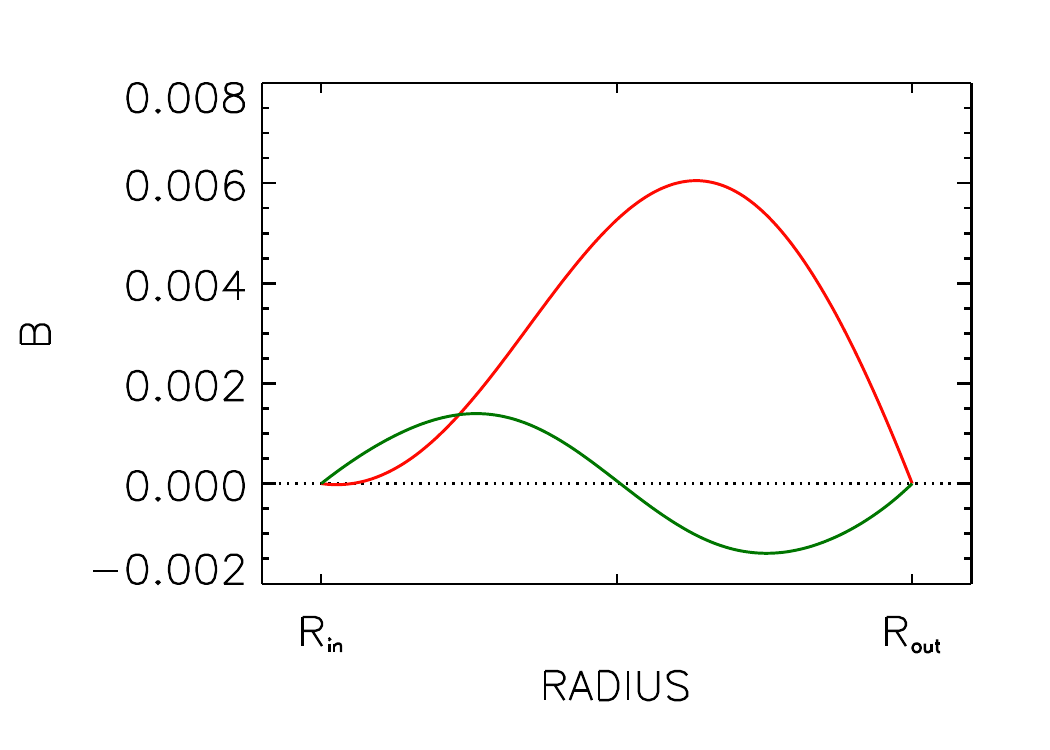}
}
}
\caption{ $\alpha\Om$ dynamo solutions in the cylindric container for the  $\alpha$ effect profiles  given in the left column.
  Middle column: proxy of the radial magnetic field $A$, right column: azimuthal magnetic field $B$. Real components in red, imaginary components in green. Vacuum boundary conditions, $C_\Om=1000$.} 
\label{fig9a}
\end{figure*}

Figure \ref{fig9} provides the smallest possible  dynamo number   in Taylor-Couette geometry (for quasi-uniform $\alpha$ effect) with $C_\alpha C_\Om \gsim 200$. The  container considered in \cite{GRE08}  provided current helicity and  $\alpha$ effect by the Tayler instability of strong toroidal fields by numerical simulations with $\Pm=15$ without density stratification but the dynamo number only reached values of  $C_\alpha C_\Om \simeq 20$ which  explains that no self-excitation appeared within the numerical experiments.
\begin{figure*}
\hbox{
\includegraphics[width=\columnwidth]{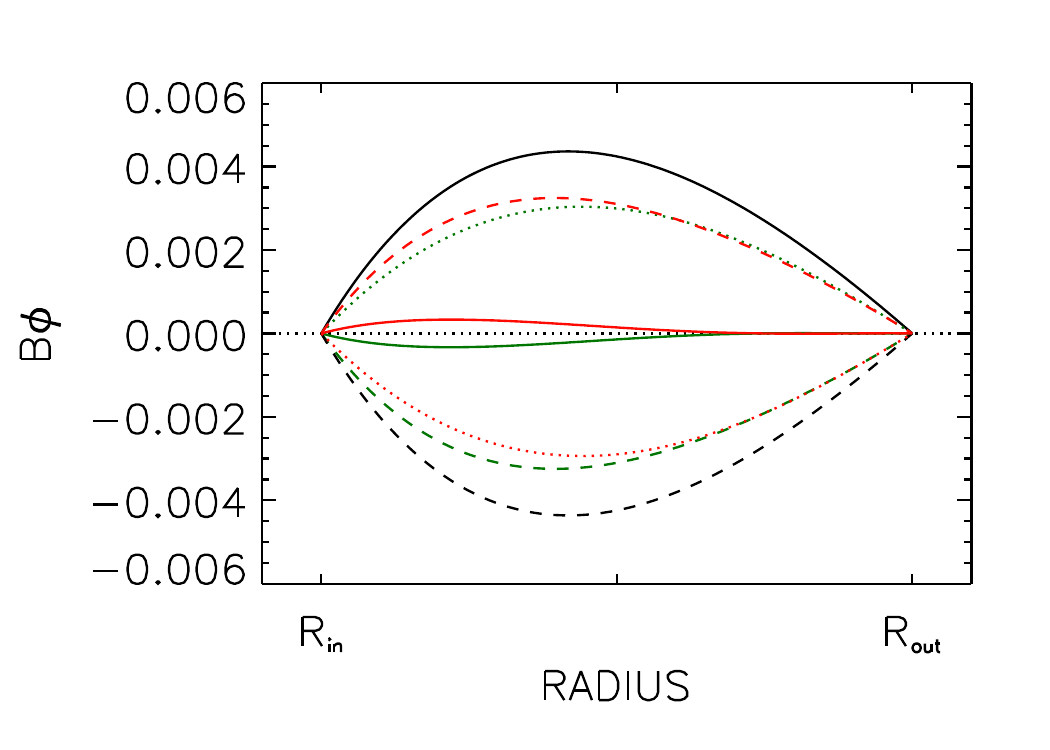}
\includegraphics[width=\columnwidth]{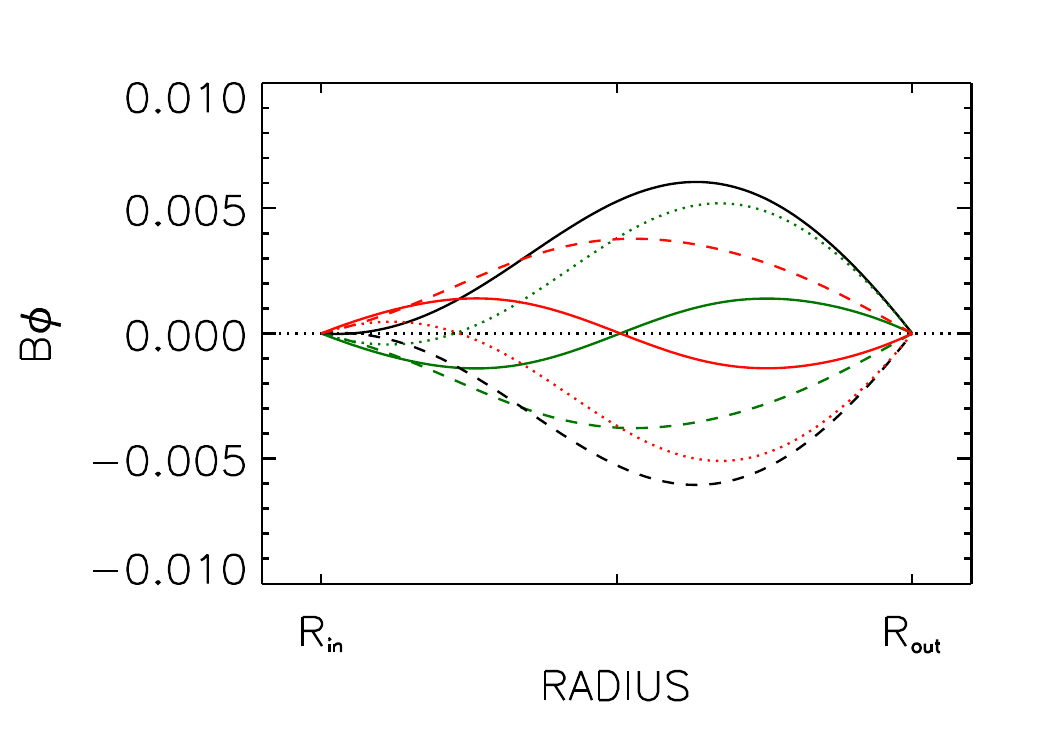}
}
\caption{Time-radius behaviour of the toroidal magnetic fields (at a fixed value of $z$) given in the right column of  Fig.   \ref{fig9a}. 
Left: artificial  quasi-uniform $\alpha$ effect  ($\omega_{\rm dyn}=-4.7$), 
Right: $\alpha$ effect with one  zero between the walls ($\omega_{\rm dyn}=-19$),
 $C_\Om=1000$. Vacuum boundary conditions. The dynamo with an $\alpha$ effect by a one-ring toroidal backgroud field finally produces two-ring fields. } 
\label{fig9c}
\end{figure*}

For  a demonstration the  typical  $\alpha$ profiles may enter the dynamo code for  $C_\Om=1000$ and  for an axial wave number which belongs to the minimum value of $C_\alpha$. Figure   \ref{fig9a} provides the results. The  magnetic potential $A$ is given in the middle column while the azimuthal field component $B$ is plotted in the right column.  The potential $A$ can be taken as  a proxy  for the radial component of the dynamo-induced field, $B_R=-{\i} K A$. The  solutions  differ in the radial patterns  of the magnetic fields from the reference solution  for the (artificial) quasi-uniform $\alpha$ profile given in the upper row.  In the zero phase of the solution only the real component (red lines) forms the magnetic field profile while  at  phase $\pi/2$ only the green line forms the field. 
The drifting solutions with a sine-type $\alpha$  always remain non-zero but at  $\pi/2$ the azimuthal magnetic field forms two magnetic rings (Fig. \ref{fig9c},  left panel). The dynamo with positive-definite $\alpha$ only produces one-cell solutions between the boundaries but the dynamo with alternating $\alpha$ --  typical for Tayler instability -- always produces two cells of opposite signature. However, the $\alpha$ effect by the Tayler instability of a background field with two antiparallel rings differs from  the characteristic $\alpha$ profile of fields with only one ring. The left panel of Fig. \ref{fig9c} shows that the dynamo for such an $\alpha$ profile only produces azimuthal fields with one ring. It cannot  be decided  how this puzzling situation influences the results of a nonlinear dynamo.

 The ratio of the maximal azimuthal field and maximal radial field in our calculations  is only of order of 30. As for $\alpha\Om$ dynamos the ratio of the two components depends on the ratio of the rotation to the $\alpha$ effect it will grow with growing $C_\Om$. The drift frequencies of the dynamos are shown in Fig. \ref{fig9d}. They are of order 0.1 the diffusion time, i.e. the TC dynamos are rather  quick oscillators. We note that for spherical convective dynamos operating in thin shear layers the ratio of the cycle time and the diffusion time is of order unity \citep{R72}.


\bibliographystyle{aa}
\bibliography{superamri}

\begin{thebibliography}{21}
\expandafter\ifx\csname natexlab\endcsname\relax\def\natexlab#1{#1}\fi

\bibitem[{{Arlt} \& {R{\"u}diger}(2011)}]{AR11}
{Arlt}, R. \& {R{\"u}diger}, G. 2011, \mnras, 412, 107

\bibitem[{{Barr{\`e}re} {et~al.}(2023){Barr{\`e}re}, {Guilet}, {Raynaud}, \&
  {Reboul-Salze}}]{B23}
{Barr{\`e}re}, P., {Guilet}, J., {Raynaud}, R., \& {Reboul-Salze}, A. 2023,
  \mnras, 526, L88

\bibitem[{{Deinzer}(1993)}]{D93}
{Deinzer}, W. 1993, in IAU Symposium, Vol. 157, The Cosmic Dynamo, ed.
  F.~{Krause}, K.~H. {Radler}, \& G.~{Rudiger}, 185

\bibitem[{{Dikpati} {et~al.}(2009){Dikpati}, {Gilman}, {Cally}, \&
  {Miesch}}]{DG09}
{Dikpati}, M., {Gilman}, P.~A., {Cally}, P.~S., \& {Miesch}, M.~S. 2009, \apj,
  692, 1421

\bibitem[{{Elstner} {et~al.}(2008){Elstner}, {Bonanno}, \&
  {R{\"u}diger}}]{EB08}
{Elstner}, D., {Bonanno}, A., \& {R{\"u}diger}, G. 2008, Astronomische
  Nachrichten, 329, 717

\bibitem[{{Gellert} {et~al.}(2008){Gellert}, {R{\"u}diger}, \&
  {Elstner}}]{GRE08}
{Gellert}, M., {R{\"u}diger}, G., \& {Elstner}, D. 2008, \aa, 479, L33

\bibitem[{{Gellert} {et~al.}(2016){Gellert}, {R{\"u}diger}, {Schultz},
  {Guseva}, \& {Hollerbach}}]{GR16}
{Gellert}, M., {R{\"u}diger}, G., {Schultz}, M., {Guseva}, A., \& {Hollerbach},
  R. 2016, \apj, 823, 99

\bibitem[{{Guseva} {et~al.}(2017){Guseva}, {Hollerbach}, {Willis}, \&
  {Avila}}]{GH17}
{Guseva}, A., {Hollerbach}, R., {Willis}, A.~P., \& {Avila}, M. 2017, \prl,
  119, 164501

\bibitem[{{Kirillov} {et~al.}(2014){Kirillov}, {Stefani}, \& {Fukumoto}}]{KS14}
{Kirillov}, O.~N., {Stefani}, F., \& {Fukumoto}, Y. 2014, Fluid Dynamics
  Research, 46, 031403

\bibitem[{{Petitdemange} {et~al.}(2024){Petitdemange}, {Marcotte}, {Gissinger},
  \& {Daniel}}]{Gissinger24}
{Petitdemange}, L., {Marcotte}, F., {Gissinger}, C., \& {Daniel}, F. 2024,
  \aap, 681, A75

\bibitem[{{Roberts}(1972)}]{R72}
{Roberts}, P.~H. 1972, Philosophical Transactions of the Royal Society of
  London Series A, 272, 663

\bibitem[{{R{\"u}diger} {et~al.}(2018){R{\"u}diger}, {Gellert}, {Hollerbach},
  {Schultz}, \& {Stefani}}]{RG18}
{R{\"u}diger}, G., {Gellert}, M., {Hollerbach}, R., {Schultz}, M., \&
  {Stefani}, F. 2018, \physrep, 741, 1

\bibitem[{{R{\"u}diger} {et~al.}(2014){R{\"u}diger}, {Gellert}, {Schultz},
  {Hollerbach}, \& {Stefani}}]{RG14}
{R{\"u}diger}, G., {Gellert}, M., {Schultz}, M., {Hollerbach}, R., \&
  {Stefani}, F. 2014, \mnras, 438, 271

\bibitem[{{R{\"u}diger} \& {Schultz}(2020)}]{RS20}
{R{\"u}diger}, G. \& {Schultz}, M. 2020, \mnras, 493, 1249

\bibitem[{{R{\"u}diger} {et~al.}(2011){R{\"u}diger}, {Schultz}, \&
  {Gellert}}]{RS11}
{R{\"u}diger}, G., {Schultz}, M., \& {Gellert}, M. 2011, Astronomische
  Nachrichten, 332, 17

\bibitem[{{R{\"u}diger} {et~al.}(2016){R{\"u}diger}, {Schultz}, {Gellert}, \&
  {Stefani}}]{RS16}
{R{\"u}diger}, G., {Schultz}, M., {Gellert}, M., \& {Stefani}, F. 2016, Physics
  of Fluids, 28, 014105

\bibitem[{{R{\"u}diger} {et~al.}(2007){R{\"u}diger}, {Schultz}, {Shalybkov}, \&
  {Hollerbach}}]{RS07}
{R{\"u}diger}, G., {Schultz}, M., {Shalybkov}, D., \& {Hollerbach}, R. 2007,
  \pre, 76, 056309

\bibitem[{{Seilmayer} {et~al.}(2012){Seilmayer}, {Stefani}, {Gundrum}, {Weier},
  {Gerbeth}, {Gellert}, \& {R{\"u}diger}}]{SS12}
{Seilmayer}, M., {Stefani}, F., {Gundrum}, T., {et~al.} 2012, Physical Review
  Letters, 108, 244501

\bibitem[{{Spruit}(2002)}]{S02}
{Spruit}, H.~C. 2002, \aa, 381, 923

\bibitem[{{Tayler}(1973)}]{T73}
{Tayler}, R.~J. 1973, \mnras, 161, 365

\bibitem[{{Vasil} {et~al.}(2024){Vasil}, {Lecoanet}, {Augustson}, {Burns},
  {Oishi}, {Brown}, {Brummell}, \& {Julien}}]{Vasil24}
{Vasil}, G.~M., {Lecoanet}, D., {Augustson}, K., {et~al.} 2024, \nat, 629, 769

\end{thebibliography}

\end{document}